\documentclass[10pt,english,aps,prb,showkeys,superscriptaddress,floatfix,twocolumn]{revtex4-2}
\usepackage[dvipsnames]{xcolor}
\usepackage{color,xcolor}
\usepackage{placeins}
\usepackage{textcomp}
\usepackage{ae,aecompl}
\usepackage[T1]{fontenc}
\usepackage[utf8]{inputenc}
\usepackage{babel}
\usepackage{amsmath}
\usepackage{amssymb}
\usepackage{graphicx}
\usepackage{wasysym}
\usepackage{booktabs}
\usepackage[normalem]{ulem}
\usepackage{lipsum}
\usepackage[unicode=true,bookmarks=true,bookmarksnumbered=false,bookmarksopen=false,breaklinks=false,pdfborder={0 0 1},backref=false,colorlinks=true,linkcolor=magenta,citecolor=blue]{hyperref}
\usepackage{contour}
\DeclareMathAlphabet{\mathpzc}{OT1}{pzc}{m}{it}
\begin{document}
\global\long\def\ket#1{\left|#1\right\rangle }%
\global\long\def\bra#1{\left\langle #1\right|}%
\global\long\def\braket#1#2{\langle#1|#2\rangle}%
\global\long\def\expectation#1#2#3{\langle#1|#2|#3\rangle}%
\global\long\def\average#1{\langle#1\rangle}%
\title{Cascade-induced high-performance nonreciprocal quantum batteries}
%

\author{Niaz Ali Khan}
\affiliation{Department of Physics, Xiamen University, Xiamen 361005, China}
\author{Dahai He}
\email{dhe@xmu.edu.cn}
\affiliation{Department of Physics, Xiamen University, Xiamen 361005, China}
\affiliation{Shenzhen Research Institute of Xiamen University, Shenzhen 518057, China}
%
\begin{abstract}
Nonreciprocal quantum batteries harness reservoir engineering for controlled energy transfer, heralding a paradigm shift for quantum energy storage. A prior study utilizing the Metelmann–Clerk formalism reported a fourfold enhancement in nonreciprocal energy accumulation over reciprocal counterparts. This fourfold enhancement, however, appears to stem from a Lindblad master equation that exhibits fundamental inconsistencies, which may systematically understate the actual nonreciprocal advantage. Our approach is rooted in the rigorously established cascaded open quantum systems formalism, which ensures both mathematical consistency and physical fidelity. Remarkably, we surpass the previously reported fourfold benchmark, attaining a regime‑independent sixteenfold steady‑state nonreciprocal energy advantage over reciprocal systems. We further uncover a dissipation-dependent battery‑to‑charger efficiency, fourfold under symmetric damping, surpassing this benchmark when the battery is less dissipative, and lower otherwise. This work establishes the cascaded formalism as a mathematically rigorous and experimentally viable foundation for high‑performance quantum energy storage.

\end{abstract}
%
\maketitle
Quantum nonreciprocity breaks time-reversal symmetry to enforce unidirectional propagation and has become a key resource in quantum technologies \cite{Kamal2011,Feng2011,Jalas2013,Sounas2017,Liang2020,Huang2021,Li2024,Zhan2025,Wang2025,Berinyuy2025,Liu2026}. It underpins directional amplification \cite{Metelmann2015,Fang2017}, quantum routing \cite{Jalas2013}, and topological protection \cite{Zhang2021}, and is realizable via reservoir engineering \cite{Poyatos1996,Metelmann2015,Fang2017,Zhang2021,Begg2024}, magneto-optics \cite{Bi2011}, nonlinear optics \cite{Khanikaev2015}, squeezing \cite{Tang2022}, non-Hermitian engineering \cite{Koutserimpas2018}, and chiral photonics \cite{Wang2025}. These mechanisms yield circulators \cite{Kamal2011}, isolators \cite{Jalas2013}, and directional amplifiers \cite{Malz2018}. Recent extensions to quantum batteries \cite{Ahmadi2024,Niaz2025,Niaz2026} demonstrate controlled unidirectional energy flow, highlighting nonreciprocity as an indispensable resource for quantum technologies.

Quantum batteries (QBs) are emerging energy storage devices that exploit quantum mechanical principles to outperform classical counterparts \cite{Alicki2013,Campaioli2017,Quach2022,Barra2019,Campaioli2024,Lu2025,Downing2024,Downing2024EPL,Song2024,Bai2020,Ahmadi2024,Andolina2025,Zhao2025,Camposeo2025,Niaz2025,Ferraro2026,Niaz2026,Hymas2026}. Their key advantages include ultrafast charging \cite{Lai2024,Andolina2025}, enhanced ergotropy (extractable work) \cite{Quach2022}, and minimal dissipation \cite{Simon2025}. Recent advances have explored diverse quantum platforms—from superconducting qubits to photonic systems—demonstrating the feasibility of quantum-enhanced charging and energy density \cite{Le2018,Ferraro2018,Rossini2020,Erdman2024,RojoFrancas2024,Hu2026}. Beyond conventional QBs, nonreciprocal designs based on reservoir engineering techniques enable unidirectional energy flow while protecting stored energy from back-action. Building on this foundation, nonreciprocal QBs \cite{Ahmadi2024,Niaz2025,Niaz2026} have recently emerged, offering directed energy flow to combat leakage and enhance efficiency. Collectively, these developments establish QBs as a foundational paradigm for next-generation energy technologies.

Recent breakthroughs in nonreciprocal quantum batteries \cite{Ahmadi2024} have demonstrated controlled energy transfer via reservoir-engineering schemes, reporting a fourfold enhancement in storage capacity over reciprocal architectures. This purported advantage, however, rests on the Metelmann-Clerk generalization \cite{Metelmann2015} of cascaded quantum systems \cite{Carmichael1991, Carmichael1993}. While conceptually innovative, that framework introduces fundamental inconsistencies into the Lindblad master equation---inconsistencies that inevitably produce unphysical predictions for battery performance. To address this shortcoming, the present work adopts the standard cascaded open quantum systems formalism \cite{Carmichael1991,Carmichael1993,Gardiner2004}. This rigorously grounded approach not only resolves the aforementioned anomalies but also delivers physically consistent and reliable results, offering a solid foundation for the realistic assessment of nonreciprocal energy-storage protocols.

\begin{figure}
\begin{centering}
\includegraphics[scale=0.85]{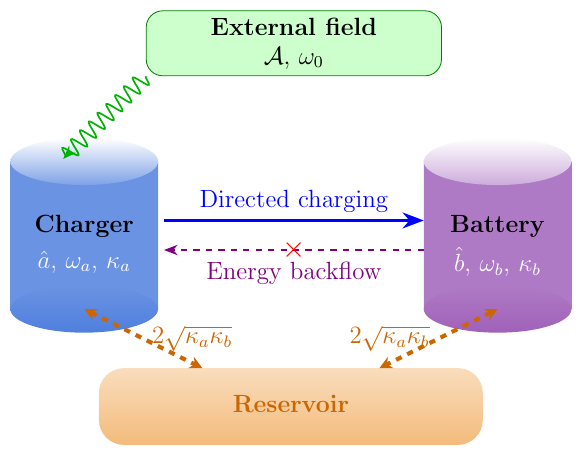}
\par\end{centering}
\caption{Schematic of the Cascaded QB model. The charger and battery are single-mode harmonic oscillators with frequencies $\omega_{a}$ and $\omega_{b}$ and local decay rates $\kappa_{a}$ and $\kappa_{b}$, respectively. The charger is driven by a coherent laser of amplitude $\mathcal{A}$ and frequency $\omega_{0}$. The reservoir engineering induces a direct coupling $\sqrt{\kappa_a \kappa_b}$ between the two systems, leading to unidirectional flow of energy from the charger to the battery.\label{fig:QuantumBatteryModel}}
\end{figure}

Our approach to nonreciprocity is fundamentally rooted in the theory of cascaded quantum systems \cite{Carmichael1991,Carmichael1993,Gardiner1993,Gardiner1985,Gardiner2004,Nha2005,Mavrogordatos2020,Wang2023,Liedl2024,Forero2025,Marcos2025}, a rigorous and experimentally established framework for engineering unidirectional coupling between quantum subsystems. This formalism offers a mathematically rigorous route to nonreciprocal interactions, making it indispensable for QBs' design and broader quantum technologies requiring directed energy or information flow. Remarkably, our findings reveal a dissipation-dependent battery-to-charger energy efficiency that scales as $4\kappa_a/\kappa_b$, a physically transparent and experimentally testable result. This efficiency achieves a fourfold enhancement under symmetric damping, surpasses this benchmark when the battery is less dissipative, and declines when the battery dominates dissipation. Most strikingly, the cascaded formalism enables us to exceed the previously reported fourfold enhancement in steady-state energy storage over reciprocal configurations, attaining a regime-independent $16$-fold enhancement. This represents a significant leap forward, establishing the cascaded approach as a platform for high-performance, practically realizable quantum batteries.
We establish a framework for a QB model comprising two single-mode harmonic oscillators, a charger (mode $a$, frequency $\omega_a$, decay rate $\kappa_a$) and a battery (mode $b$, frequency $\omega_b$, decay rate $\kappa_b$) coupled to a shared reservoir, as illustrated in Fig.~\ref{fig:QuantumBatteryModel}. The charger is driven by a classical field of amplitude $\mathcal{A}$ and frequency $\omega_0$, whereas the battery is initialized in its ground state. We emphasize that the shared-reservoir coupling considered here constitutes a convenient implementation of nonreciprocal dynamics, as the reservoir modes mediate the interactions between the two harmonic oscillators. Our nonreciprocity approach is rooted in cascaded quantum systems theory \cite{Carmichael1991,Carmichael1993,Gardiner1993,Gardiner1985,Gardiner2004}, which enables unidirectional coupling between quantum subsystems. By tracing out the reservoir degrees of freedom under the Born-Markov approximation, the reduced density matrix of the oscillators $\rho$ in the rotating frame evolves according to the master equation
\begin{align}
\dot{\rho} = \frac{1}{i\hbar}\left[\mathcal{H},\rho\right] + \mathcal{D}_{\mathcal{M}}(\rho),
\label{eq:LindbladEquation0}
\end{align}
where $\mathcal{M} = \sqrt{2\kappa_a} a + \sqrt{2\kappa_b} b$ is the joint collapse operator. The first term in Eq.~\eqref{eq:LindbladEquation0} describes the coherent Hamiltonian dynamics, whereas the second term, $\mathcal{D}_{\mathcal{M}}(\rho)$, is the Lindblad superoperator, defined as
\begin{align}
\mathcal{D}_{\mathcal{M}} (\rho)=& \mathcal{M}\rho \mathcal{M}^{\dagger}-\frac{1}{2} \left\{ \mathcal{M}^{\dagger}\mathcal{M},\rho\right\},
\label{eq:SuperOperator0}
\end{align}
Applying the rotating-wave approximation with respect to the driving field, the Hamiltonian $\mathcal{H}$ for the engineered reservoir QB system takes the form
\begin{align}
\mathcal{H}=& \hbar \Delta_{a} a^{\dagger} a + \hbar\Delta_{b} b^{\dagger} b+ i \hbar \sqrt{\kappa_a \kappa_b}(a^{\dagger}b-ab^{\dagger}) \nonumber\\ &
+\hbar\mathcal{A}(e^{i\Delta_a t} a + e^{-i\Delta_a t} a^{\dagger}),
\label{eq:Hamiltonian}
\end{align}
where $\Delta_x=\omega_0-\omega_x$ ($x \in \{a, b\}$) is the laser–charger/battery detuning. We take $\omega = \omega_a = \omega_b = \omega_0$ (resonance) without loss of generality. The Hamiltonian $\mathcal{H}$ captures the distinct couplings of the charger and battery to the reservoir modes. Remarkably, as a consequence of reservoir engineering, the two systems acquire a direct coupling of strength $\sqrt{\kappa_a \kappa_b}$, leading to unidirectional flow of energy.
\FloatBarrier
The nonreciprocal behavior of the cascade open quantum system arises from the interplay between coherent and reservoir-induced dissipative interactions. Employing the cascaded open quantum systems formalism \cite{Carmichael1991,Carmichael1993,Gardiner1993,Gardiner1985}, the dynamics are governed by the Lindblad master equation [Eq.~\eqref{eq:LindbladEquation0}] for the density matrix $\rho$, which may equivalently be expressed as
\begin{align}
\dot{\rho}=& \frac{1}{i\hbar}\left[\mathcal{H},\rho\right]+2\sum_{i\in\{a,b\}}\kappa_{i}\mathcal{L}_{i}^{i}(\rho)+ 2\sqrt{\kappa_a\kappa_b}\left[\mathcal{L}_{a}^{b}(\rho) \nonumber \right. \\ & \left.  +\mathcal{L}_{b}^{a}(\rho)\right],
\label{eq:LindbladEquation}
\end{align}
These processes are encapsulated in the Lindblad superoperators $\mathcal{L}_{x}^{y}(\rho)$ with $x,y\in \{a,b\}$, defined as
\begin{align}
\mathcal{L}_{x}^{y}(\rho)=& x\rho y^{\dagger}-\frac{1}{2} \left\{ x^{\dagger}y,\rho\right\}.
\label{eq:SuperOperater}
\end{align}
where $\mathcal{L}_{i}^{i}(\rho)$ with $i\in \{a,b\}$ models the local dissipation of systems. The last terms, $\mathcal{L}_{a}^{b}(\rho)$ and $\mathcal{L}_{b}^{a}(\rho)$, constitute the crucial cascaded coupling, which captures the collective dissipation arising from coupling both subsystems to a common engineered reservoir and serves as the source of nonreciprocity.

The Metelmann–Clerk generalization of cascaded quantum systems theory \cite{Metelmann2015} facilitates a broad class of directional interactions. Considering a pair of cavity modes each coupled to a reservoir, they engineered a nonreciprocal interaction for nonreciprocal transmission or amplification of incident signals. Upon elimination of this reservoir, an effective dissipative interaction emerges between the cavities. Assuming a Markovian reservoir, the dynamics are governed by a Lindblad master equation for the reduced density matrix $\rho$ of the two cavity modes, subject to the nonreciprocal condition by setting $J=i\mu\Gamma/2$.
\begin{align}
\dot{\rho}=& \frac{1}{i\hbar}\left[\mathcal{H}_0,\rho\right]+\sum_{i\in\{a,b\}}\kappa_{i}\mathcal{L}_{i}^{i}(\rho)+\Gamma\mathcal{L}_{z}(\rho),
\label{eq:LindbladEquationMetalmann}
\end{align}
where $\mathcal{H}_0=\hbar \Delta_{a} a^{\dagger} a + \hbar\Delta_{b} b^{\dagger} b+ \hbar (Ja^{\dagger}b+J^*ab^{\dagger})+\hbar\mathcal{A}(e^{i\Delta_a t} a + e^{-i\Delta_a t} a^{\dagger})$ is the total Hamiltonian in the rotating-wave approximation and $\mathcal{L}_{z}(\rho)$ captures the collective dissipation arising from the coupling of both subsystems to a common engineered reservoir. This is governed by a collective jump operator of the form $z=u_a a+u_{b}b$, with $u_{a}\,\,(u_{b})$ representing the complex coupling coefficient between the charger (battery) and the reservoir \cite{Ahmadi2024, Metelmann2015}. 
Setting $\Gamma=2\sqrt{\kappa_a\kappa_b}$ and comparing Eq.~(\ref{eq:LindbladEquation}) with Eq.~(\ref{eq:LindbladEquationMetalmann}), we observe that the first two terms are equivalent and display analogous behavior within the model. However, the collective dissipation terms differ. The last term in the Metelmann–Clerk approach is
\begin{align}
\mathcal{L}_{z}(\rho)=& \sum_{i\in\{a,b\}}\mathcal{L}_{i}^{i}(\rho)+\mathcal{L}_{a}^{b}(\rho)+ \mathcal{L}_{b}^{a}(\rho) ,
\label{eq:LindbladEquationCollective}
\end{align}
where the extra local Lindblad superoperator terms $\mathcal{L}_{i}^{i}(\rho)$ $(i=a,b)$, which appear in the collective dissipation, lead to anomalous effects.  
Apparently, the generalized method that extends the fundamental structure of cascaded quantum systems is expressed as:
\begin{align}
\dot{\rho}=& \frac{1}{i\hbar}\left[\mathcal{H}_0,\rho\right]+\sum_{i\in\{a,b\}}(\kappa_{i}-\Gamma \left|u_{i}\right|^{2})\mathcal{L}_{i}(\rho)+\Gamma\mathcal{L}_{z}(\rho).\label{eq:LindbladMasterEquation}
\end{align}
This method can be employed in the design of nonreciprocal cavity-based photonic devices, wherein a given coherent interaction is matched with its corresponding dissipative counterpart. The Supplemental Material \cite{SupplementalMaterial} provides details on nonreciprocal QBs using the general method for constructing nonreciprocal systems by matching a coherent interaction with its dissipative counterpart.

To evaluate the dynamics of nonreciprocal QBs, we employ two key figures of merit: the energy storage capacity $\mathcal{E}(t)$ and the ergotropy $\mathcal{W}(t)$. These respectively quantify the energy transfer efficiency from the charger to the battery and the maximum extractable work from the battery. The energy stored in the battery over charging time $t$ is given by
\begin{align}
\mathcal{E}_b(t)=& \text{Tr}\left[ \mathcal{H}_{b} \rho_{b}(t)\right] =\hbar\omega\average{b^{\dagger}b},
\label{eq:EnergyStored}
\end{align}
where $\omega$ is the frequency of the harmonic oscillator battery, and $\text{Tr}$ denotes the trace over the battery's reduced state, $\rho_{b}(t)$. Thus, $\mathcal{E}_b(t)$ measures the mean energy stored in the battery at time $t$. To quantify the useful work capacity, we employ ergotropy, a central concept in quantum thermodynamics that measures the maximum work extractable from a quantum system via a cyclic unitary process \cite{Farina2019},
\begin{align}
\mathcal{W}_{b}(t)=& \text{Tr}\left[ \mathcal{H}_{b} \rho_{b}(t)\right]-\underset{U_{b}}{\min}\text{Tr}\left[\mathcal{H}_{b}U_{b}\rho_{b}(t)U_{b}^{\dagger}\right],
\label{eq:Ergotropy0}
\end{align}
where the minimization in Eq.~\eqref{eq:Ergotropy0} is over all local unitaries $U_{b}$ acting on the battery. The second term in Eq.~\eqref{eq:Ergotropy0} corresponds to the passive energy of the system—the portion of its energy that is inaccessible under any unitary operation $U_{b}$. Since the system is Gaussian, its ergotropy can be expressed in terms of operator expectation values of the mode operators as follows \cite{Downing2024}:
\begin{align}
\mathcal{W}_{b}(t)=& \mathcal{E}_b(t)-E_b (t),
\label{eq:Ergotropy}
\end{align}
where $E_b (t)$ is the energy of the corresponding passive state of the battery, from which no work can be extracted cyclically via unitary operations \cite{Downing2024EPL}. The energy of the passive state can have the form:
\begin{align}
E_b(t)=& \frac{\hbar\omega}{2} \left(\sqrt{\mathcal{D}}-1\right),
\label{eq:passiveEnergy}
\end{align}
where $\mathcal{D}$ is a dimensionless composite parameter that captures the battery's first and second moments, taking the form
\begin{align}
\mathcal{D}=\left[1+2\left(\average{b^{\dagger}b}-\left|\average{b}\right|^{2}\right) \right]^2-4\left|\average{bb}-\average{b}^2\right|^{2}.
\label{eq:D-factor}
\end{align}

To characterize the nonreciprocal QB, we derive the equations of motion for the mode operators. Under the resonance frequency condition, the Lindblad master equation yields the following first-order moments for the charger and battery modes:
\begin{align}
\dot{\average{a}}= -\kappa_{a}\average{a}-i\mathcal{A}, \quad
\dot{\average{b}}= -\kappa_{b}\average{b}-2 \sqrt{\kappa_a\kappa_b}\average{a},
\label{eq:1stMomenta}
\end{align}
Similarly, the second-order moments of the mode operators in nonreciprocal settings are analytically expressed as
\begin{align}
\average{\dot{a^{\dagger} a}}=& -2\kappa_{a}\average{a^{\dagger} a}+i\mathcal{A}\left(\average{a}-\average{a^{\dagger}} \right),\nonumber\\
\average{\dot{b^{\dagger}b}}=& -2\kappa_{b}\left\langle b^{\dagger}b\right\rangle -2\sqrt{\kappa_a\kappa_b}\left(\average{a^{\dagger}b}+\average{ab^{\dagger}}\right),\nonumber\\
\average{\dot{a^{\dagger}b}}=&-(\kappa_{a}+\kappa_{b})\average{a^{\dagger}b}+i\mathcal{A}\average{b}-2\sqrt{\kappa_a\kappa_b}\average{a^{\dagger} a},\nonumber\\
\average{\dot{a^2}}=& -2\kappa_{a}\average{a^2}-2i\mathcal{A}\average{a},\nonumber\\
\average{\dot{b^2}}=& -2\kappa_{b}\average{b^2}-4\sqrt{\kappa_a\kappa_b}\average{ab},\nonumber\\
\average{\dot{ab}}=&-(\kappa_{a}+\kappa_{b})\average{ab}-i\mathcal{A}\average{b}-2\sqrt{\kappa_a\kappa_b}\average{a^2}.
\label{eq:2ndMoment}
\end{align}
It is evident from Eqs.~\eqref{eq:1stMomenta} and \eqref{eq:2ndMoment} that the engineered reservoir enforces a unidirectional interaction: the battery dynamics are driven by the charger, but not vice versa. This nonreciprocity establishes a unidirectional energy channel, preferentially directing energy from charger to battery while suppressing backflow. The reservoir mediates a nonlocal damping force coupling the two modes, and the asymmetry in the coupling coefficients decouples the charger's dynamics from the battery's, thereby breaking time-reversal symmetry.
\FloatBarrier
%
In what follows, solving the coupled differential equations with initial conditions $\average{\mathcal{O}}|_{t=0}=0$, where $\average{\mathcal{O}}$ denotes the expectation values of the mode operators, yields the battery energy
\begin{align}
\mathcal{E}^{\rm nr}_b(t)=\frac{4\hbar\omega\kappa_a\kappa_b\mathcal{A}^2}{(\kappa_a-\kappa_b)^2} \left[\frac{1-e^{-\kappa_a t}}{\kappa_a}-\frac{1-e^{-\kappa_b t}}{\kappa_b}\right]^2,
\label{eq:nBatteryEnergy}
\end{align}
and the charger energy, $\mathcal{E}^{nr}_a(t)$
\begin{align}
\mathcal{E}^{\rm nr}_a(t)=\frac{\hbar\omega\mathcal{A}^2}{\kappa_a^2} \left(1-e^{-\kappa_a t}\right)^2.
\label{eq:nChargerEnergy}
\end{align}
For a QB system coupled to a zero-temperature Markovian bath (vacuum reservoir), all second-order correlators factorize into products of first-order moments: $\average{a^{\dagger} a}=\left|\average{a}\right|^{2}$, $\average{aa}=\average{a}^2$, $\average{b^{\dagger}b}=\left|\average{b}\right|^{2}$, $\average{bb}=\average{b}^2$. Hence, the dimensionless parameter $\mathcal{D}$ becomes unity, giving zero passive state energy, and the stored energy is fully extractable: $\mathcal{E}_{b}^{\rm nr}(t)=\mathcal{W}_{b}^{\rm nr}(t)$. This ergotropy results from nonreciprocal coupling, which suppresses backflow and preserves coherence. In contrast, under a squeezed thermal reservoir, factorization breaks down, and ergotropy no longer equals the total stored energy \cite{Niaz2026}.
%
\FloatBarrier
Under symmetric dissipative coupling ($\kappa_{a}=\kappa_{b}=\kappa$), the nonreciprocal energy $\mathcal{E}^{\rm nr}_{b}(t)$ simplifies to:
\begin{align}
\mathcal{E}^{\rm nr}_b(t)=\frac{4 \hbar \omega\mathcal{A}^2}{\kappa^2} \left[1-(1+\kappa t)e^{-\kappa t}\right]^2.
\label{eq:nBatteryEnergyIdenticalDamping}
\end{align}
Our analysis reveals that the charging performance (stored energy and ergotropy) of the QB critically depends on the damping rates and the amplitude of the driving field. The energy dynamics exhibit an inverse square dependence on the damping parameter, implying that minimizing the damping rate enhances the maximum energy storage capacity. However, this strategy directly conflicts with the need for rapid charging. As established in Eq.~(\ref{eq:Hamiltonian}), the charger-battery coupling strength is proportional to $\sqrt{\kappa_a \kappa_b}=\kappa$; thus, reducing the damping rates to increase capacity inevitably weakens the coupling, leading to prohibitively long charging times. This reveals a fundamental trade-off between energy capacity and charging power. Consequently, a strategy focused solely on minimizing the damping rates to maximize energy capacity is self-defeating, as it catastrophically suppresses the charging rate. Achieving practical utility therefore requires a careful balance, sacrificing some maximum energy storage to secure a sufficiently strong coupling and ensure viable charging performance. Optimal design must be guided by this fundamental trade-off between high capacity and efficient energy transfer. In the steady-state limit $(t\rightarrow \infty)$, Eq.~(\ref{eq:nBatteryEnergyIdenticalDamping}) converges to $4\omega\mathcal{A}^2/\kappa^2$.
%
\begin{figure}
\begin{centering}
\includegraphics[scale=0.35]{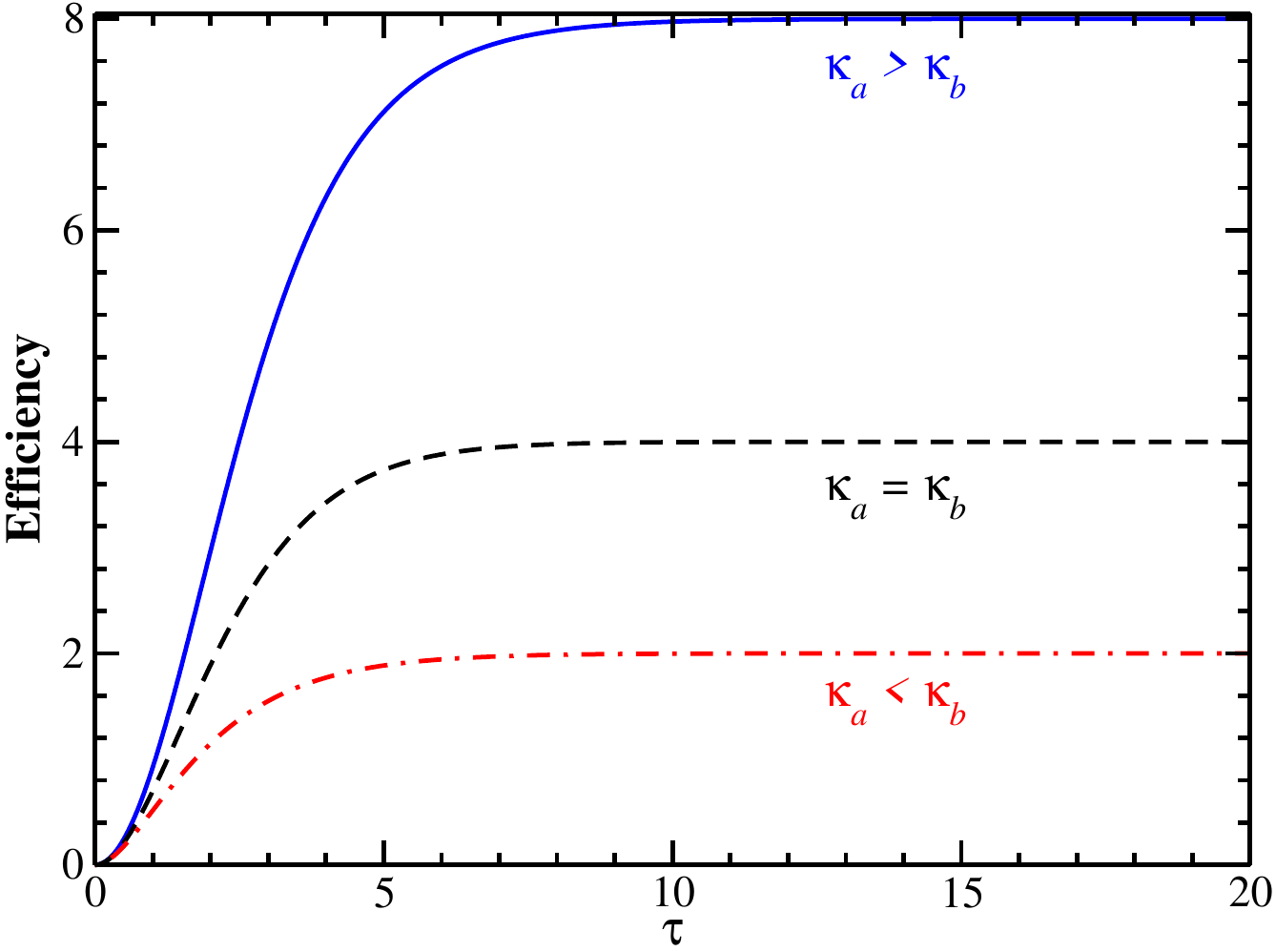}
\par\end{centering}
\caption{Energy efficiency (battery-to-charger energy ratio) of the nonreciprocal QB system as a function of rescaled time $\tau$. The system's parameters are $\kappa_a =0.02\omega, \,\, \kappa_b = 0.01\omega$ (blue bold); $\kappa_a = \kappa_b$ (black dashed); and $\kappa_a =0.01\omega,\,\, \kappa_b = 0.02\omega$ (red dotted-dashed).\label{fig:BCfficiency}}
\end{figure}
\begin{figure}
\begin{centering}
\includegraphics[scale=0.35]{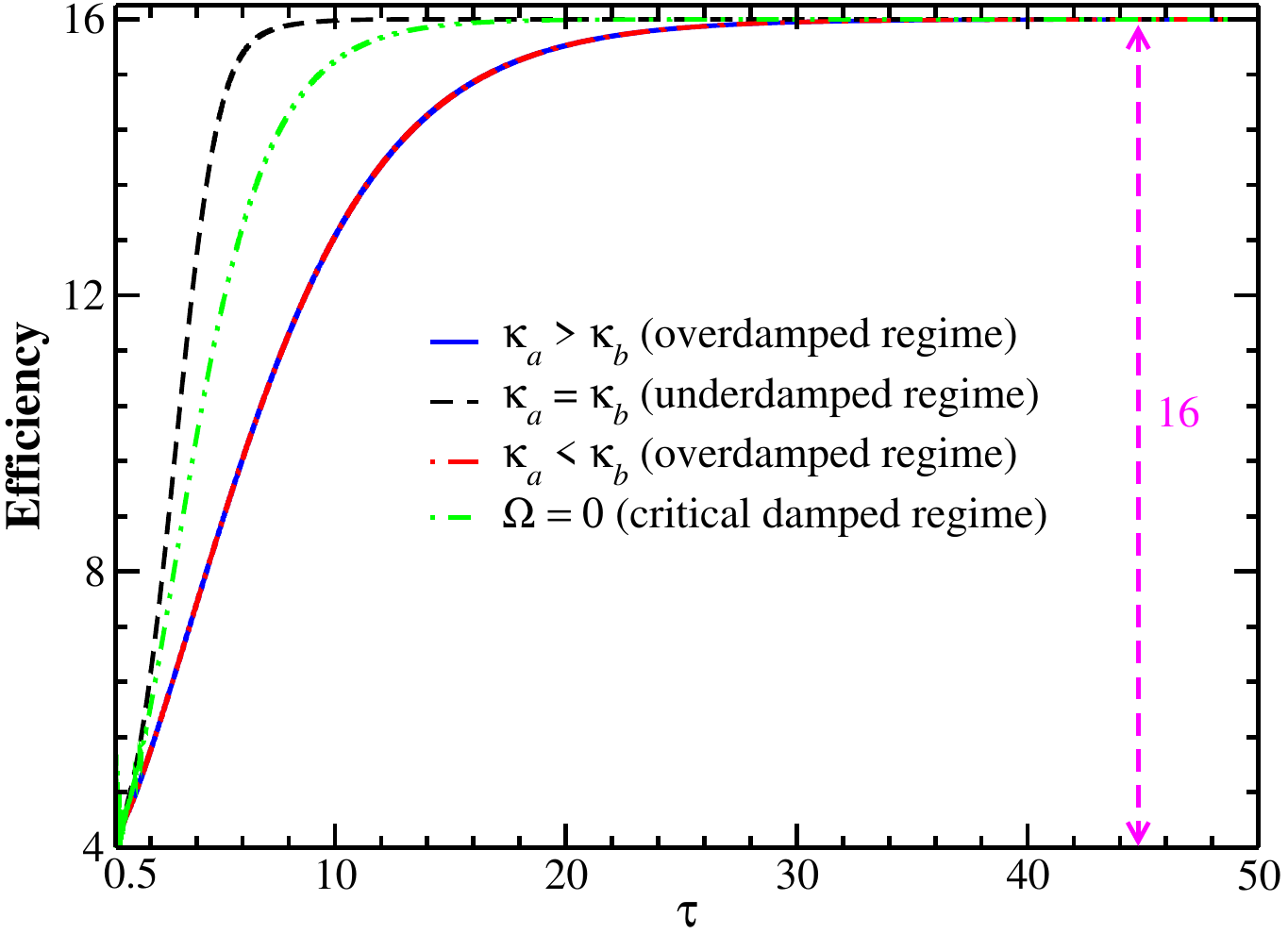}
\par\end{centering}
\caption{Energy efficiency (nonreciprocal-to-reciprocal stored energy ratio) versus rescaled time $\tau$ of the QB system.  The system's parameters are $\kappa_a =0.2\omega, \,\, \kappa_b = 0.01\omega$ (blue bold); $\kappa_a = \kappa_b=0.01\omega$ (black dashed); $\kappa_a =0.01\omega,\,\,\kappa_b = 0.2\omega$ (red dotted-dashed); and $\Omega\approx0$ (green dotted-dashed). To enable a fair comparison, the coupling strength is set to $\sqrt{\kappa_a \kappa_b}$ for reciprocal configurations.\label{fig:nBrBefficiency}}
\end{figure}

A more stringent metric is the efficiency, which measures the ratio of the battery's energy to that of the charger. For instance, the performance of the nonreciprocal charging scheme is assessed by comparing these two energies. Accordingly, we define the energy efficiency $\eta(t)$ as
\begin{align}
\eta(t) = \frac{\mathcal{E}^{\rm{nr}}_{b}(t)}{\mathcal{E}^{\rm{nr}}_{a}(t)}.\label{eq:EfficiencyBA}
\end{align}
Importantly, as reported in \cite{Ahmadi2024}, the steady-state battery-to-charger energy efficiency scales as $4(1+\kappa_b/\Gamma)^{-2}$, attaining the fourfold enhancement only when $\kappa_b/\Gamma\rightarrow0$, \emph{i.e.,} either $\kappa_b\rightarrow0$ (ideal battery) or $\Gamma\rightarrow\infty$ (infinite coupling). Moreover, under identical damping $\Gamma=\kappa_b$, the system exhibits no genuine nonreciprocal advantage, as the stored battery energy equals the charger energy. By contrast, within the standard cascaded formalism adopted in the present work, the steady-state energy efficiency converges to the markedly different expression
\begin{align}
\eta(t\rightarrow\infty) = 4\frac{\kappa_a}{\kappa_b},\label{eq:EfficiencySteadyState}
\end{align}
which explicitly depends on the charger-to-battery damping ratio rather than on an idealized limiting condition. This result underscores a fundamental discrepancy between the two theoretical frameworks: whereas the Metelmann–Clerk generalization predicts an unbounded advantage contingent upon extreme parameter regimes, the rigorously grounded cascaded approach yields a finite, physically transparent efficiency that is inherently constrained by the system's dissipative parameters.

Figure~\ref{fig:BCfficiency} illustrates the energy efficiency (battery-to-charger energy ratio) of the nonreciprocal quantum battery under vacuum reservoir engineering. The efficiency evolves monotonically with rescaled time and asymptotically approaches the steady-state limit given in Eq.~\eqref{eq:EfficiencySteadyState}, which corresponds to the maximum extractable energy and ergotropy. Crucially, this physically realistic behavior emerges only when the standard cascaded formalism is employed, further motivating its adoption over the Metelmann–Clerk framework.

The performance of the nonreciprocal charging scheme can also be evaluated by comparing the battery's energy with that of a conventional reciprocal battery. The reciprocal regime is obtained by eliminating the shared reservoir, and its stored energy $\mathcal{E}_b^{\rm r}(t)$ is given by
\begin{align}
\mathcal{E}^{\rm r}_{b}(t)=&  \frac{\hbar\omega\mathcal{A}^2\mathcal{C}}{\kappa_{a}\kappa_{b}\left(\mathcal{C}+1\right)^{2}} \left[1+\mathcal{G}(t)e^{-\kappa_{ab}t}-\mathcal{F}(t) e^{-\frac{\kappa_{ab}t}{2}}\right],
\label{eq:ReciprocalEnergy}
\end{align}
with
\begin{align}
\mathcal{G}(t)=& \frac{(\Omega -\kappa_{ab})^2}{2\Omega ^2}\cosh{\Omega t}+\frac{\kappa_{ab}}{\Omega }e^{\Omega t}-\frac{2\kappa_{a}\kappa_{b}}{\Omega ^2}\left(\mathcal{C}+1\right),\nonumber\\
\mathcal{F}(t)=& 2 \cosh{\frac{\Omega  t}{2}}+\frac{2\kappa_{ab}}{\Omega }\sinh{\frac{\Omega  t}{2}}
\label{eq:rEnergyTerms}
\end{align}
where $\mathcal{C} = J^2 / \kappa_a \kappa_b$ is the is the charger-battery cooperativity cooperativity, $\kappa_{ab} = \kappa_a + \kappa_b$ is the total local dissipation rate, and $\Omega^2  = |\kappa_a - \kappa_b|^2 - |2J|^2$. The nature of $\Omega$ demarcates the dynamical regime: an imaginary $\Omega $ yields underdamped dynamics, while a real $\Omega$ signifies the overdamped regime.

The nonreciprocal quantum batteries reported in \cite{Ahmadi2024} achieve a fourfold energy storage advantage over reciprocal architectures. Crucially, however, this enhancement is realized only in the limiting cases $\kappa_b\rightarrow0\,(\text{with} \,\,\Gamma>0)$ or $\Gamma\rightarrow\infty \,(\text{with} \,\,\kappa_b>0)$, corresponding respectively to dissipative cooperativity $\mathcal{C}\rightarrow\infty$ or $\mathcal{C}\rightarrow0$, where $\kappa_b$ is the battery dissipation rate and $\Gamma$ is the reservoir coupling strength. The former limit describes an ideal, lossless battery, while the latter corresponds to complete decoupling between charger and battery---neither of which represents a physically realistic operating regime. For any finite, nonzero values of $\kappa$ or $\Gamma$, the purported enhancement falls strictly below the fourfold threshold, thereby diminishing the practical significance of the claimed advantage. This parametric fragility, together with the foundational inconsistencies inherent in the Metelmann-Clerk generalization, motivates the present work's adoption of the standard cascaded open-quantum-systems formalism. Figure~\ref{fig:nBrBefficiency} presents the nonreciprocal-to-reciprocal energy efficiency of the quantum battery model under this rigorously grounded framework. Remarkably, the efficiency saturates at a value of $16$, corresponding to a sixteenfold nonreciprocal advantage that substantially surpasses the fourfold enhancement previously reported in \cite{Ahmadi2024}. This enhancement, achieved monotonically with respect to scaled time, establishes a new benchmark for quantum battery charging performance. Dynamically, the underdamped regime approaches steady state more rapidly than its critical or overdamped counterparts, owing to more efficient transient energy exchange. Crucially, and in marked contrast to prior work—wherein the overdamped regime severely degraded performance---our asymptotic efficiency proves entirely independent of the damping regime. This regime independence overcomes a fundamental limitation of earlier nonreciprocal charging schemes and, more broadly, demonstrates the robustness and predictive power of the standard cascaded formalism across the full dissipative-parameter landscape.
Remarkably, this work outperforms all previous Metelmann–Clerk–based research across multiple domains, including  quantum-limited amplification \cite{Metelmann2014},  three-mode bosonic system \cite{Pocklington2023}, chiral quantum optics  \cite{Forero2025}, superoptimal charging of QBs \cite{Ahmadi2025}, and thermally shielded work potential \cite{Niaz2026}.

In conclusion, we have identified a fundamental limitation in Metelmann–Clerk–based nonreciprocal QB designs. Previous claims of a fourfold enhancement in nonreciprocal energy accumulation were found to be contingent upon unphysical parameter regimes, stemming from mathematical inconsistencies in the Lindblad master equations introduced by the Metelmann–Clerk generalization of cascaded quantum systems. To overcome this limitation, we adopted the standard cascaded open quantum systems technique and thereby uncovered a dissipation-dependent efficiency (battery-to-charger energy ratio) scaling as $4\kappa_a/\kappa_b$. This gave a fourfold enhancement under symmetric damping, exceeding this benchmark when the battery was less dissipative and lower otherwise. Most notably, we surpassed the previously reported fourfold benchmark, thereby achieving a regime-independent sixteenfold steady-state nonreciprocal energy advantage over reciprocal systems. This work thus established the cascaded formalism as a mathematically rigorous and experimentally viable foundation for high-performance quantum energy storage.

\textit{Acknowledgments}---This work was financially supported by the National Natural Science Foundation of China (Grant No. 12475039), and Guangdong Basic and Applied Basic Research Foundation (Grant No. 2025A1515010350).
\bibliographystyle{apsrev4-2.bst}
\bibliography{QBs}

\end{document}